\documentclass[conference,letterpaper]{IEEEtran}
\usepackage{makeidx}
\usepackage[british,UKenglish,USenglish,english,american]{babel}
\usepackage[utf8]{inputenc}
\usepackage[dvips]{graphicx}
\usepackage{graphicx,url}
\usepackage{amssymb}%fancyhdr,tabularx,epsfig}
\usepackage{psfrag,amsmath}
\usepackage{multirow}
\usepackage[Lenny]{fncychap}
\usepackage{fancyhdr}
\usepackage{makeidx}
\usepackage{epsfig}
\usepackage{graphicx,color}

\usepackage{array}
\usepackage{arydshln}
\usepackage[all]{xy}
\usepackage{float}
\usepackage{hyperref}
\usepackage{multicol}
\usepackage[margin=10pt,font=scriptsize,labelfont=bf,labelsep=endash]{caption}

\usepackage[T1]{fontenc}
\usepackage{ifthen}
\usepackage{cite}
\usepackage{stfloats} %pacote para colocar a tabela/figura em qualquer parte de um texto de duas colunas 
\usepackage{blindtext} %tabelas com duas colunas de texto 
\newtheorem{coro}{Corollary}

\newtheorem{defi}{Definition}
\newtheorem{teo}{Theorem}
\newtheorem{lema}{Lemma}
\newtheorem{obs}{Remark	}

\newtheorem{prop}{Proposition}
\newtheorem{rema}{Remark}
\newcommand{\Fn}{\mathbb{F}_2^n}
\newcommand{\Fm}{\mathbb{F}_2^m}
\newcommand{\Fnm}{\mathbb{F}_2^{n+m}}

\newcommand{\supp}{\mathrm{supp}}
\newcommand{\argmin}{\mathrm{arg \ min}}
\newcommand{\bx}{\mathbf{x}}
\newcommand{\by}{\mathbf{y}}
\newcommand{\bz}{\mathbf{z}}
\newcommand{\bzero}{\mathbf{0}}
\newcommand{\bc}{\mathbf{c}}
\newcommand{\be}{\mathbf{e}}
\newcommand{\ba}{\mathbf{a}}
\newcommand{\bb}{\mathbf{b}}

\begin{document}
	\title{Metrics which turn tilings into binary perfect codes} 
	
	% %%% Single author, or several authors with same affiliation:
	% \author{%
	%   \IEEEauthorblockN{Stefan M.~Moser}
	%   \IEEEauthorblockA{ETH Zürich\\
	%                     ISI (D-ITET)\\
	%                     CH-8092 Zürich, Switzerland\\
	%                     Email: moser@isi.ee.ethz.ch}
	% }

	%%% Several authors with up to three affiliations:
	\author{%
		\IEEEauthorblockN{Gabriella Akemi Miyamoto}
		\IEEEauthorblockA{IMECC - 
						University of Campinas\\
				gabriellaakemimiyamoto@gmail.com}
		\and
		\IEEEauthorblockN{Marcelo Firer}
		\IEEEauthorblockA{IMECC - University of Campinas\\
				 mfirer@ime.unicamp.br}
	}

\maketitle

\begin{abstract}
In this work, we consider tilings of the Hamming cube and look for metrics which turn the tilings into a perfect code. We consider the family of metrics which are determined by a weight and are compatible with the  support of vectors (TS-metrics). We determine which of the tilings with small tiles or high rank can be a perfect code for some TS-metric and we characterize all such metrics. Finally, we show some procedures to obtain  new perfect  codes (relatively to TS-metrics) out of existing ones.  
\end{abstract}

\section{Introduction}
The study of perfect codes is an important topic in coding theory, since it satisfies an optimality condition: the coincidence between the packing and covering radii. Finding perfect codes is a difficult issue.  For the Hamming metric, there is a complete characterization of its parameters, which are the parameters of a trivial code, a Hamming code  \cite{hamming} and Golay codes \cite{golay1}. In Lee metric, the situation is more unclear (see \cite{golomb2, golomb,post}). Besides Van Lint's good survey of perfect codes from 1975 \cite{Lint}, a more recent survey was made by Olof in 2008 \cite{olof}. 

\medskip
The concept of perfect code, that is, the coincidence between the packing and covering radii, can  naturally be stated for any discrete metric $d(\cdot ,\cdot )$, so we stress the metric in question by saying it is a $d$-perfect code. 

\medskip

In this work we are concerned with perfect codes when considering a particular but reasonable family of metrics on $\mathbb{F}_2^n $, called TS-metrics, which stands for \emph{invariant by \textbf{T}ranslations } and \emph{respecting \textbf{S}upport}. These are metrics that are defined by a weight and respects the support of vectors %, in the sense that the natural ordering of the supports (under the inclusion) is compatible with the ordering of the weights 
(details on Section \ref{subsec:TS}). These two properties are quite natural to be asked when considering linear binary error correcting codes. It is worth noting that such metrics admit a syndrome decoding algorithm and, under some circumstances, this algorithm may have a significant gain in reducing the table of coset leaders (see \cite[Section 4.1]{lucianofelix}).
\vspace{2pt}

There are two large families of TS-metrics, namely the poset metrics and the combinatorial metrics, introduced respectively by R. Brualdi at. al in \cite[1995]{brualdi} and E. Gabidulin in \cite[1973]{gabidulin}.
In this more general setting, the only family of metrics in which perfect codes were studied are the so called poset metrics. A recent account of it can be found in \cite[Chapter 6.3.1]{livro}. 

\medskip

Our approach has two steps that are simple to explain.

\begin{enumerate}
	\item If we have a tiling of the Hamming cube and each tile is a  ball for a given metric $d$, then the center of the balls constitute a $d$-perfect code. So, we  consider known tilings of the Hamming cube and ask which of these tilings can be a metric ball of a TS-metric. For those that satisfy this condition, we try to classify (up to equivalence) all such metrics. 
	\item The second step asks to construct new perfect codes out of existing ones. To be more precise, given a $d_1$-perfect code on $\mathbb{F}_2^{n_1}$ and a $d_2$-perfect code on $\mathbb{F}_2^{n_2}$ we try to find a metric $d$ that turns the concatenation of the two codes into a $d$-perfect code in $\mathbb{F}_2^{n_1+n_2}$. 
\end{enumerate}	 

The main source of existing tilings of the Hamming cube, and of ways to construct tilings out of existing ones  is \cite{vardy}, our main reference for this text. In that work, the authors present a  complete classification of small tilings (for tiles with up to eight elements) and tilings with tiles of high rank. To be more clear, to each tiling presented in \cite{vardy} we first determine if there exists a metric for which it is a perfect code. After that, for those which admit such a metric, we classify all the metrics that do it. This is the starting point of our first approach. They also show that tilings are invariant by concatenation, and we use it for our second step. 

Due to lack of space, proofs are omitted, but can be found in a complete version of this work in \cite{akemiarxiv}.

\section{Preliminaries}

Throughout this paper, let   $\mathbb{F}_2^n$ be the $n$-dimensional vector space over $\mathbb{F}_2$,  $[n]=\{1,...,n\}$ and $\supp(\bx):=\{i\in [n] ; x_i\not= 0\}$ the  \textbf{support} of $\bx\in \mathbb{F}_2^n$. We let $\omega_H$ and $d_H$ denote the Hamming weight and Hamming metric, respectively.

\subsection{TS-metrics}\label{subsec:TS}

The Hamming metric has two important properties, expressed in the next two definitions.
\vspace{2pt}

\begin{defi}
	A metric $d:\mathbb{F}_2^n \times \mathbb{F}_2^n \to \mathbb{R}$ is said to be \textbf{translation-invariant} if 
	\[d(\bx+\bz,\by+\bz)=d(\bx,\by)\]
	for every $\bx ,\by,\bz\in \mathbb{F}_2^n$.
\end{defi}

It is well known and worth noting that a metric is translation-invariant \emph{iff} it is defined by a weight \footnote{A function $\omega:\mathbb{F}_2^n\to \mathbb{R}$ is a \textbf{weight} if it satisfies the following axioms: $(1)$  $\omega(\bx)\geq 0$ for every $\bx$; $(2)$  $\omega(\bx)=0$ if, and only if, $\bx=0$; $(3)$ $\omega(\bx+\by)\leq \omega(\bx)+\omega(\by)$. A weight determines a metric by defining $d(\bx,\by)=\omega(\bx-\by)$. }.

\vspace{2pt}
\begin{defi}
	A weight function $\omega$ is said to \textbf{respect the support} of vectors if 
	$\supp(\bx)\subseteq \supp(\by) \Longrightarrow \omega(\bx)\leq \omega(\by).$
\end{defi}

\vspace{2pt}
	A \textbf{translation-support metric} (\textbf{TS-metric}) is a metric that is  translation-invariant and which respects the support of vectors.
	
	Being translation-invariant is a key property for decoding linear codes, since syndrome decoding depends exclusively on this property.  
	
	Respecting the support of vectors is a property that is crucial in coding theory (for binary codes), once it means that making extra errors cannot lead to a better situation, in the sense that making an error on the coordinate $i$ in a message cannot be worse than making two errors, one on the coordinate $i$ and the other on $j$. %By the translation-invariant property, it is enough to take $0$ as the center of the balls. 

\medskip
We present now the two principal families of TS-metrics which will be explored in this work.
\vspace{2pt}

\subsubsection{Poset Metric}

The poset metrics were introduced by Brualdi et al. in \cite{brualdi}. 

Let $P=([n],\preceq)$ be a partially ordered set (\textbf{poset}).  An \textbf{ideal} in a poset $P=([n],\preceq)$ is a nonempty subset $I\subseteq [n]$ such that, for $a\in I$ and $b\in [n]$, if $b\preceq  a$ then $b\in I$. We denote by $\langle I\rangle$ the ideal generated by $I\subseteq [n]$. %The subscript $P$ may be omitted when there is no risk of ambiguity. 
An element $a$ of an ideal $I\subset [n]$ is called a \textbf{maximal} element of $I$ if $a\preceq  b$ for some $b\in I$ implies $b=a$. %The set of all maximal elements of an ideal $I$ is denoted by $\mathcal{M}_{P}(I)$. Notice that, given an ideal $I\subseteq [n]$, we have that $\langle \mathcal{M}_{P}(I) \rangle_{P}=I$ and $I$ is minimal with this property.  
We say that $b$ \textbf{covers} $a$ if $a\preceq b$, $a\not= b$ and there is no extra element $c\in [n]$ such that $a\preceq c\preceq  b$. 
\vspace{2pt}

\begin{defi}
The \textbf{$P$-weight} of a vector $\bx\in\mathbb{F}_2^n$ is defined by
\[\omega_{P}(\bx):=|\langle \supp(\bx)\rangle|,\]
where $|A|$ is the cardinality of $A$.
\end{defi}
\vspace{2pt}

The $P$-weight clearly respects support, since $A\subset B$ implies $\langle A\rangle \subset \langle B \rangle$. The \textbf{$P$-distance} in $\mathbb{F}_2^n$ is the metric induced by $\omega_{P}$: $d_{P}(\bx,\by):=\omega_{P}(\bx-\by)$.

\vspace{2pt}
\subsubsection{Combinatorial Metric}

The combinatorial metrics were introduced by Gabidulin in \cite{gabidulin}. 

Let $\mathbb{P}_n=\{A ; A\subset[n]\}$ be the power set of $[n]$. We say that a family $\mathcal{A}\subset \mathbb{P}_n$ is a \textbf{covering} of a set $X\subset[n]$ if $X\displaystyle\subset \bigcup_{A\in\mathcal{A}}A$. 

If $\mathcal{F}$ is a covering of $[n]$, then the \textbf{$\mathcal{F}$-combinatorial weight} of $\bx=(x_1,...,x_n)\in \mathbb{F}_2^n$ is the integer-valued map $\omega_{\mathcal{F}}$ defined by
\[\omega_{\mathcal{F}}(\bx):=\mbox{min}\{|\mathcal{A}| ; \mathcal{A}\subset \mathcal{F}, \mathcal{A} \text{ is a  covering of } \supp(\bx) \}.\]

The distance defined as $d_{\mathcal{F}}(\bx,\by):=\omega_{\mathcal{F}}(\bx-\by)$ is called \textbf{$\mathcal{F}$-combinatorial metric}.

We denote by ${\mathcal{TS}(n)}$, ${\mathcal{P}(n)}$ and ${\mathcal{C}(n)}$ the sets of all TS-metrics, poset and combinatorial, respectively. It is worth to note that ${\mathcal{P}(n)}, {\mathcal{C}(n)}\subset {\mathcal{TS}(n)}$.

\subsection{Perfect codes}
Given a metric  $d$ on $\mathbb{F}_2^n$, the \textbf{ball of radius $r$ and center $\bx$} is 
		$B_d(\bx,r)=\{ \by\in \mathbb{F}_2^n ; d(\bx,\by)\leq r\}.$ A code $C\subseteq \mathbb{F}_2^n$ is a $(d,r)$-\textbf{perfect code} if $\displaystyle \bigcup_{\bc\in C} B_d(\bc,r) =\mathbb{F}_2^n$ and $B_d(\bc,r) \cap B_d(\bc^\prime,r) =\emptyset, \forall \bc,\bc^\prime\in C, \bc\neq\bc^\prime$.
	
We approach now the first of our key definitions. 
\vspace{2pt}

\begin{defi}
Given a subset $S\subseteq \Fn$, we say that $S$ is a \textbf{TS-ball} if $S$ is a ball for some TS-metric, that is, $S=B_d(\bx,r)$, for some $\bx\in \Fn$, $r>0$ and $d\in \mathcal{TS}(n)$. If $C$ is a $(d,r)$-perfect code for some $d\in\mathcal{TS}(n)$ we say that it is a \textbf{TS-perfect code}. In case the radius $r$ is not taken into consideration, we say $C$ is $d$-perfect.
\end{defi}

\subsection{Tiles, tilings  and polyhedrominoes}

%\subsection{Tilings of $\mathbb{F}_2^n$}

We are interested in building perfect codes out of tilings of the Hamming cube, so we need some basic definitions about tilings and polyhedrominoes.

 A \textbf{path} $\gamma$ in $\mathbb{F}_2^n$, with initial point $\bx$ and final point $\by$, is a sequence $\gamma:\bx_{0},\bx_{1},...,\bx_{t}$, where $d_{H}\left(  \bx_{i},\bx_{i+1}\right)  =1$, $\bx=\bx_{0}$ and $\by=\bx_{t}$. The \textbf{length} of $\gamma$ is defined by $\left\vert \gamma\right\vert =t $. A path $\gamma$ is called a \textbf{geodesic path} if it is a path of minimum length between the initial and final points.  A path $\gamma$ from $\bx$ to $\by$ is a geodesic path  if, and only if, $d_{H}\left(  \bx,\by\right)=\left\vert \gamma\right\vert $.

%A subset $V\subset\mathbb{F}_{2}^{n}$ is said to be \textbf{connected} if given $v,w\in V$ exists a path $\gamma$ connecting $v$ to $w$ with $\gamma\subset V$. $V$ is a \textbf{convex set} if exists a geodesic $\gamma$ connecting $v$ to $w$ with $\gamma\subset V$. Obviously, any convex set is also connected, but the converse is not true.
%
%A subset $V\subset\mathbb{F}_{2}^{n}$ is \textbf{strongly convex at a point $x\in V$}  if given $v\in V$, \textbf{every} geodesic $\gamma$ connecting $x$ to $v$ satisfies $\gamma\subset V$.

% If $D$ is a strongly convex set related to the center $x$, then $D$ is a \textbf{convex polyhedromino}.
\vspace{2pt}

\begin{defi}
A set $D\subseteq \mathbb{F}_2^n$ is a \textbf{polyhedromino} if for all $\bx,\by\in D$ there is a (possibly not unique) geodesic path $\gamma\subset D$ connecting $\bx$ to $\by$. %A polyhedromino is \textbf{convex} if, for all $u,v\in D$ we have $\gamma\subset D$ for every geodesic path $\gamma$ connecting $u$ to $v$ . 
	
\end{defi}
\vspace{2pt}

%In graph theory, given a graph $G$ and a subgraph $H$, an $H$-tiling in $G$ is a collection of vertex-disjoint copies of $H$ in $G$, that is, $G$ is tiled (covered) by translated copies of $H$. For finite fields, the idea is the same and it is presented below.
We adopt the definition of tiling given in \cite{branko}, since it makes evident its relation to perfect codes. It is not difficult to see that this is equivalent to the definition adopted in our main reference \cite{vardy}.

\vspace{2pt}
\begin{defi}\label{tiling1}
	A \textbf{tiling} of $\mathbb{F}_2^n $ is a pair $(D,C)$, where $D,C \subseteq \mathbb{F}_2^n$ and $C$ is a subset such that 
	\[\displaystyle \bigcup_{\bc\in C} \bc+D =\mathbb{F}_2^n \ \ \mbox{and} \ \ (\bc+D) \cap(\bc^\prime+D) =\emptyset,\]
	$\forall \bc,\bc^\prime\in C, \bc\not=\bc^\prime$.

\end{defi}
 \vspace{2pt}
 
Despite the fact that the role of $D$ and $C$ are interchangeable, we shall call $D$ as a \textbf{tile} and $C$ the code of the tiling, since this is the role it will play in the context of coding theory. In the case where $D$ is a polyhedromino, we say $(D,C)$ is a \textbf{poly-tiling} of $\mathbb{F}_2^n $. %If $D$ is a convex polyhedromino, then $(D,C)$ is a \textbf{convex tiling}.
 
Since we are working with translation-invariant metrics, it is always possible to translate all the elements of both $D$ and $C$ in order to have  $\bzero\in D$ and $\bzero\in C$. Then, throughout this paper, w.l.o.g., we may assume that $\bzero\in D$ and $\bzero\in C$.

Notice that we are considering only translated copies of $D$, which is very reasonable in the context of TS-metrics, since in this case all the translated copies of the tile are isometric. Also, as we shall see, it is also reasonable the use of polyhedrominoes to tile $\Fn$.

Tilings and perfect codes are two distinct research areas. Tilings are frequently studied in the context of graph theory and notice that a particular case of graph is the Hamming graph. Next proposition establishes a connection between tilings and perfect codes. 

\vspace{2pt}
\begin{prop}\label{prop:poly}
Given $(D,C)$ a tiling of $\mathbb{F}_2^n$, suppose that $D=B_d(\bzero,r)$ for some  $d\in \mathcal{TS}(n)$. Then:

\begin{enumerate}
	\item $D$ is a polyhedromino;
	\item $C$ is a $(d,r)$-perfect code. 
\end{enumerate} 
\end{prop}

\begin{IEEEproof}
The proof follows directly from the definitions and it is omitted due to lack of space.	
\end{IEEEproof}	
In case the conditions of the proposition holds, we say that the tiling $(D,C)$ \textbf{determines a TS-perfect code}.

	\begin{table*}[b]
	\centering
	\begin{tabular}{|c|c|c|c|c|}
		\hline
		Tile & Rank & Elements & Radius & Non trivial relations of the Poset \\
		\hline
		$D_1^3$ & $3$ & ${\bzero}, \be_1, \be_2, \be_3, \be_1+\be_2, \be_1+\be_3,\be_2+\be_3, \be_1+\be_2+\be_3 $ & 3 & $P_1:1\preceq 2\preceq 3$, \\\hline
		$D_1^7$ & $7$ & $\bzero, \be_1, \be_2, \be_3, \be_4, \be_5, \be_6, \be_7$ & 1 & $P_2:$ only trivial relations\\ \hline
	\end{tabular}
	\caption{Tiles of type 2}
	\label{tilesball}
\end{table*}

\begin{table*}[b]
	\centering
	\begin{tabular}{|c|c|c|c|c|}
		\hline
		Tile & Rank & Elements & Radius & Combinatorial metric\\
		\hline
		$D_1^4$ & 4 & ${\bzero}, \be_1, \be_2, \be_3, \be_4, \be_1+\be_2, \be_1+\be_3,\be_1+\be_4$ & 1 & $\mathcal{F}_1=\{\{1,2\},\{1,3\},\{1,4\}\}$\\ \hline
		$D_2^4$ & 4 & ${\bzero}, \be_1, \be_2, \be_3, \be_4, \be_1+\be_2, \be_1+\be_3,\be_2+\be_3$ & 1 & $\mathcal{F}_2=\{\{1,2\},\{1,3\},\{2,3\},\{4\}\}$ \\ \hline
		$D_1^5$ & 5 & ${\bzero}, \be_1, \be_2, \be_3, \be_4, \be_5, \be_1+\be_4, \be_1+\be_5$ & 1 & $\mathcal{F}_3=\{\{1,4\},\{1,5\},\{2\},\{3\}\}$\\\hline
		$D_1^6$ & 6 & ${\bzero}, \be_1, \be_2, \be_3, \be_4, \be_5, \be_6, \be_1+\be_2$ & 1 & $\mathcal{F}_4=\{\{1,2\}, \{3\}, \{4\}, \{5\}, \{6\}\}$ \\\hline
	\end{tabular}
	\caption{Tiles of type 2}
	\label{tilesball2}
\end{table*}

A trivial (and not interesting) way of obtaining a poly-tiling is to consider $I\subset [n]$, $D_I=\{\bx =(x_1,\ldots ,x_n);x_i=0, i\in I\}$ and $C_I=\{\bx =(x_1,\ldots ,x_n);x_i=0, i\in [n]\setminus I\}$. It is also trivial to see that given a tiling $(D,C)$, we have that $|D|\cdot |C|=|\mathbb{F}_2^n|$.

%Our main reference \cite{vardy} adopts a different definition for tilings. 
%\vspace{2pt}
% 
%\begin{defi}\cite{vardy}\label{tiling2}
%The pair $(D,C)$ is a tiling of $\mathbb{F}_2^n$ if $D+C=\mathbb{F}_2^n$ and $2D\cap 2C=\{\bzero\}$, where both $D$ and $C$ contain the element ${\bzero}$.
%\end{defi}
%
%
%It is not difficult to see that the definitions  \ref{tiling1} and \ref{tiling2} are equivalent. 

\section{Obtaining perfect codes out of tilings}
The starting point of this section is the work \cite{vardy}, where  tilings of $\mathbb{F}_2^n$ with ``small'' tiles were classified, where a tile $D$ is called ``small'' if  $|D|\leq 8$. Since a tiling $(D,C)$ satisfies $|D|\cdot|C|=2^n$,  we must have $|D|$ equals $1,2,4$ or $8$. %The first case is rather trivial and will not be treated in this work.

In Section \ref{subsec:tiles} we obtain all small tilings $(D,C)$ presented in \cite{vardy} and determine each of those $C$ is a TS-perfect code;  In Section \ref{sec:large} we give necessary and sufficient conditions for a tiling of large rank presented in \cite{vardy} to determine a TS-perfect code; In Section \ref{extensaotile}, given a perfect code $(D,C)$ with respect to a metric $d$ and with $D\subset\mathbb{F}_2^s$, we present a systematic way to extend $d$ into a metric $d^\ast$ on $\Fn$ which turns the extension of $(D,C)$ to $\Fn$ to be a perfect code; Finally, in Section \ref{equivmetrics} we classify all TS-metrics that turn $D$ into a ball or equivalently, turn C into a TS-perfect code.    

%   we characterize the properties of a ball in a TS-metric. On the second part, we list all the small tilings described in \cite{vardy} and determine which one is a ball (or not) for some TS-metric. In \cite{vardy}, they characterize all tiles of $\mathbb{F}_2^n$ up to size 8. In the case where the tiling cardinality is equal to 8 elements, they separate the tiles by rank, which variate from 3 to 7. \\

\subsection{Classifying small tiles that determine TS-perfect codes}\label{subsec:tiles}

We denote by $\be_i\in\mathbb{F}_2^n$ the vector with $\supp (\be_i)=\{i\}$.
\vspace{2pt}

\begin{prop}\label{prop1}
Let $B=B_d({\bzero},r)\subseteq\mathbb{F}_{2}^{n}$ be a TS-ball with $2$ or $4$ elements. Then, $B$ is one of the following:
\begin{align*}
	B_1 & = \left\{ \bzero,\be_{i}\right\},\\
	B_{2}  & = \left\{  \bzero,\be_{i},\be_{j},\be_{k}\right\}  , i,j,k \text{ distincts }\\
	B_{3}  & =\left\{  \bzero,\be_{i},\be_{j},\be_{i}+\be_{j}\right\}, i\neq j.
	\end{align*}
\end{prop}

\begin{IEEEproof}
	The tiles listed are all polyhedrominoes of this size hence, by Proposition \ref{prop:poly} these are all the possible candidates. They are all realized by a poset metric, determined, respectively, by the non-trivial sets of relations: $\{i\preceq l;\forall l \neq i  \}  $, $\{ t\preceq l; \forall t\in \{i,j,k\}, l\in [n]\setminus \{i,j,k\}   \}  $ and $\{ t\preceq l; \forall t\in \{i,j\}, l\in [n]\setminus \{i,j\}   \}  $.
		\end{IEEEproof}

The \textbf{rank of} $V\subset \mathbb{F}_2^n$ is the dimension of the vector subspace generated by $V$, i.e., $rank(V)=dim\langle V \rangle$. Given a tiling the rank of $(D,C)$ is $rank(D,C)=rank(D)$.

In \cite{vardy} there is a complete classification  of tilings of $\mathbb{F}_2^n$. To obtain the first result of this section, Proposition \ref{nottile}, there are two steps: first to reduce the list of  classification in \cite{vardy} by considering equivalents tiles that can be obtained by a simple permutation of the coordinates, obtaining $15$ equivalence classes. We remark that, if $D=B_d(\bzero,r)$ is a ball for some TS-metric and   $\bx\in B_d(\bzero,r)$, then $\by\in B_d(\bzero,r)$ for all  $\by\in\Fn$ such that $\supp(\by)\subseteq \supp(\bx)$. This simple remark makes possible to eliminate $9$ of those tiles, which do not satisfy this property.  As an example, let $D=\{{\bzero}, \be_1, \be_2, \be_3, \be_4, \be_1+\be_3, \be_1+\be_4,\be_1+\be_3+\be_4\}$ be a tile. Note that  $\supp(\be_3+\be_4)\subset \supp(\be_1+\be_3+\be_4)$ and $\be_3+\be_4\notin D$, then we have $\omega(\be_3+\be_4)\geq \omega(\be_1+\be_3+\be_4)$. Then, by the remark above, $D$ cannot be a ball for any TS-metric.

The remaining $6$ tiles are presented in Tables \ref{tilesball} and \ref{tilesball2}. They are denoted by $D_{j}^s$, where $s$ is the rank of the tile and $j$ is a counting index. Hence we have the following: 

\medskip

\begin{prop}\label{nottile}
	If a tile is not equivalent to a tile presented in  Table \ref{tilesball} or \ref{tilesball2}, there is no TS-metric that turns it into a ball.
\end{prop}

To show that the remaining tiles give rise to a TS-perfect code, we need to find a TS-metric which turns them into a metric ball. The proof of the next theorem, Theorem \ref{teo:TS}, is actually the last column of the tables, where we present a poset metric (for Table \ref{tilesball}) or a combinatorial metric (for Table \ref{tilesball2}) that turns the tile into a ball.

\vspace{2pt}

\begin{teo}\label{teo:TS}
For each tile $D$ in Tables \ref{tilesball} and \ref{tilesball2} there exists a TS-metric on $\mathbb{F}_2^s$ for which $D$ is a ball, where $s=rank(D)$.
\end{teo}

\begin{IEEEproof}
The proof consists in exhibiting a TS-metric for each case. The last column of Tables \ref{tilesball} and \ref{tilesball2} exhibits an appropriate TS-metric for which $D$ is a ball. Each case should be directly verified. 
\end{IEEEproof}
\vspace{2pt}

\begin{obs}
	The tiles $D$ listed in Tables \ref{tilesball} and \ref{tilesball2} are considered as subsets of $\mathbb{F}_2^s$, where $s=rank(D)$. In Section \ref{extensaotile} we show a process used to extend them to $\Fn$, $n\geq s$.
\end{obs}

\subsection{Classifying tiles with large rank that determine TS-perfect codes}\label{sec:large}

In \cite{vardy}, the authors proved that a set $D_n(\bx)= \{\be_i;i\in [n]  \} \cup \{ \bzero, \bx\}  $ , for some $\bx\in\Fn$ with $\omega_H(\bx)\geq 2$ is a tile if, and only if, $\omega_H(\bx)\notin \{n-1,n-2 \}$. We shall determine a necessary and sufficient condition for it to define a TS-perfect code.
\vspace{2pt}

\begin{prop}
Suppose that $(D_n(\bx),C_n(\bx))$ is a tiling of $\Fn$. Then, there is a TS-metric that turns it into a perfect code if, and only if, $\omega_H(\bx)=2$. 
\end{prop}

\begin{IEEEproof}
	If $\omega_H(\bx)>2$, then $D_n(\bx)$ cannot be a ball in a metric that respects support, since in this case there would be some subset $A\subset\supp (\bx)$ with $1<|A|<\omega_H(\bx)$ and the vector $\bx_A$ defined by $\supp (\bx_A)=A$ is not contained in $D_n(\bx)$. For $\omega_H(\bx)=2$, we have that $\bx=\be_j+\be_k$, for some $j\neq k$ and we define $\mathcal{F} = \{  \{i\} ; i\in [n] \}  \cup \{\{ j,k \} \}$ and we have that $D_n(\bx)=B_{d_\mathcal{F}}(\bzero,1)$ and, by Proposition \ref{prop:poly} we have that $(D_n(\bx),C_n(\bx))$ is a $d_\mathcal{F}$-perfect code.
		\end{IEEEproof}

\subsection{Extending tilings from $\mathbb{F}_2^s$ to $\Fn$}\label{extensaotile}

Given $\ba=(a_1,a_2,...,a_n)\in \mathbb{F}_2^n$ and $\bb=(b_1,b_2,...,b_m)\in \mathbb{F}_2^m$, $\ba\mid \bb=(a_1,a_2,...,a_n,b_1,b_2,...,b_m)$ and $A\mid B=\{\ba\mid \bb; \ \ba\in A, \ \bb\in B\}$.

\medskip
In the previous section we considered tilings $(D,C)$ of $\mathbb{F}_2^s$ where $s=rank (D)$. Since $\mathbb{F}_2^s$ can be seen as a linear subspace of $\Fn$ for $n\geq s$, we can extend this to a tiling $(D^\ast ,C^\ast)$ of $\Fn$. We denote $\bzero_l$ the null element in $\mathbb{F}_2^l$ and let $D^\ast =D\mid \bzero_{n-s}$ and $C^\ast=C \mid \mathbb{F}_2^{n-s}$. As can be found in  \cite{vardy} we have that $(D^\ast ,C^\ast)$ is a tiling of $\Fn$. We remark that $|D^\ast |=|D|$.

If $(D,C)$ is a tiling of $\mathbb{F}_2^s$ and $d\in\mathcal{TS}(s)$ turns $D$ into a metric ball $B_d(\bzero,r)$ in $\mathbb{F}_2^s$ (or equivalently, turns $C$ into a $d$-perfect code), we wish to extend $d$ to a metric $d^\ast$ which turns $D^\ast$ into a metric ball $B_{d^\ast}(\bzero,r^\prime)$ in $\Fn$. 

\medskip
\begin{teo}\label{teo:exttiles}
	Given $D=B_{d}(\bzero,r)$, $d\in \mathcal{TS}(s)$, there is  $d^\ast\in \mathcal{TS}(n)$ such that $D^{\ast}=B_{d^\ast}(\bzero,r)$.
\end{teo}

\begin{IEEEproof}
Given a  weight $\omega$ on $\mathbb{F}_2^s$, let $M(\omega) = \max\{\omega(\bx);\bx\in\mathbb{F}_2^s  \} $. We define, for $\bx\in\Fn$, $n\geq s$
\[
\omega_{n,s} (\bx)= \begin{cases} \omega (\bx) \text{ if } \supp(\bx)\subset [s]\\
M(\omega )+1 \text{ otherwise}\end{cases}.
\]

It is not difficult to see that $\omega_{{n,s}} (\bx)$ is a weight. Let $d$ and $d_{n,s}$ be the metrics determined by $\omega$ and $\omega_{{n,s}}$ respectively. It is not difficult to prove that  $d$ respects the support of vectors if, and only if,  $d_{n,s}$ does it. Moreover,
	\[
B_{d_{n,s}}(\bzero,r) =	B_d(\bzero,r)\mid \{ \bzero_{n-s} \}
	\]	
for every $r\leq M(\omega )$. So, if $(D,C)$ determines a perfect code, so does $(D^\ast ,C^\ast)$.
\end{IEEEproof}

\vspace{2pt}
\begin{rema}
In the two cases considered in Table \ref{tilesball}, where the metrics were determined by a poset $P$ over $[s]$, it is possible to   extend it to a metric defined by a poset $P^\ast$ over $[n]$ as follows:
	 $P_1^{\ast}$ is defined by the (non-trivial) relations  $1\preceq 2\preceq 3$ and $3\preceq i$ for all $i\geq 4$. The poset  $P_2^{\ast}$ is defined by the (non-trivial) relations $i\preceq j$ for all $ i\leq 7 < j$. These are actually the minimal poset metrics which extend the original ones and it is not difficult to classify all the poset extensions that do it.
	 	 
	 For the cases in Table \ref{tilesball2}, the extension follows by directly applying Theorem \ref{teo:exttiles}.
\end{rema}

\bigskip

\subsection{Classifying the TS-metrics which turn a tiling into a perfect code}\label{equivmetrics}

If $(D,C)$ determines a perfect code, there is $d\in\mathcal{TS}(n)$ that turns $D$ into a metric ball. Actually, there are infinitely many such metrics, so when we wish to classify all such metrics, we mean up to an equivalence relation. The most natural equivalence relation in the context of coding theory is to say that two metrics on $\Fn$ are equivalent if they determine the same minimum distance decoding for every code $C\subset \Fn$ and every received message $\bx\in\Fn$. To be more precise:

\begin{defi}\label{def:equiv}
Two metrics (or distances) $d_1$ and $d_2$ defined over $\mathbb{F}_2^n$ are \textit{decoding equivalent}, denoted by $d_1\sim d_2$, if 
\[\argmin\{d_1(\bx,\bc): \bc\in C\} = \argmin\{d_2(\bx,\bc): \bc\in C\},\]
for any code $C\subseteq \mathbb{F}_2^n$ and any $\bx\in \mathbb{F}_2^n$.
\end{defi}
\vspace{3pt}

It is not difficult to see that $d_1\sim d_2$ if, and only if, $d_1 (\bx,\by)<d_1 (\bx,\bz)\iff d_2 (\bx,\by)<d_2 (\bx,\bz)$, for all $\bx,\by,\bz\in\mathbb{F}_2^n$. Details about this equivalence relation can be found in \cite{rafael} and \cite{rafael2}. 

%In \cite{rafael}, the authors presented a manner to verify which are the equivalent metrics to a given one. In this section, we use their result to show which are the equivalent metrics to the metrics that made the tiles (in Tables \ref{tilesball} and \ref{tilesball2}) a ball.

Let $M\subset \mathbb{F}_2^{N}\times \mathbb{F}_2^{N}$, $N=2^n$ be a distance matrix where $m_{\bx,\by}=d(\bx,\by)$ and $d\in\mathcal{TS}(n)$. Our goal is to determine necessary and sufficient conditions (on the matrix $M$) to determine a TS-metric that turns a tiling $(D,C)$ into a perfect code. This is what is done in the next theorem.

%Let $D$ be a tile that is determined by a TS-metric (those in Tables \ref{tilesball} and \ref{tilesball2} and let  $d$ be the metric determined in Proposition \ref{exttiles} which turns $D$ into a metric ball of radius $r$ in $\mathbb{F}_2^s$. Consider a matrix  
\vspace{2pt}

\begin{teo}
		Let $(D,C)$ be a tiling of $\Fn$. Let $d$ be a TS-metric for which $D=B_d(\bzero,r)$.	Let
		 $M =(m_{\bx,\by})\subset \mathbb{F}_2^{N}\times \mathbb{F}_2^{N}$ be a $N\times N$ matrix, with $N=2^n$, satisfying the following conditions:
		\begin{enumerate}
			\item[C1)]\label{c1} $m_{\bx,\bzero} = d(\bx,\bzero)$ for $\bx\in D$.
			\item[C2)] $m_{\bx,\bzero}>r$ for $\bx\notin D$.
			\item[C3)] $m_{\bx,\by}=m_{\by-\bx,\bzero}$ for all $\bx,\by\in\mathbb{F}_2^N$.
			
		\end{enumerate}
	
	Then, the following holds:
\begin{enumerate}
	\item[i)] The matrix $M$ defines a distance which is decoding-equivalent  to a metric $d_M$  that is a translation-invariant metric. 
	\item[ii)] The tile $D$ is a metric ball of the metric $d_M$, to be more precise, $D=B_{d_M}(\bzero,r)$.
	\item[iii)] It is possible to choose the values of $m_{\bx,\by}>r$ for $\bx\notin D$ in such a way that the metric $d_M\in \mathcal{TS}(N)$; 
	\item[iv)] Any TS-metric $d'$ which turns $D$ into a metric ball is equivalent to a metric described by a matrix $M$ satisfying conditions C1, C2, C3.

\end{enumerate}

\end{teo}

\begin{IEEEproof} We briefly sketch the main steps in the proof.
	The existence of a metric follows from the symmetry of the matrix (since on a binary space $\bx-\by=\by-\bx$) and the fact that on a finite space any distance is equivalent to a metric (see \cite[Chapter 1.1]{deza2009}). The translation invariance follows from the fact that the first row determines all the others. Second item follows from the fact that $m_{\bx,\bzero}\leq r$ if, and only if, $ \bx\in D$. The third is done constructively and the last one follows from the algorithm presented in \cite{rafael} to obtained a reduced form of a metric.
\end{IEEEproof}

%We notice that, in the binary case $\mathbb{F}_2$ we have $m_{x,y}=m_{y,x}$, that is, $M$ is symmetric and also, $M$ is determined by its first row, because due to the fact that $d_M$ is a translation invariant metric, the others rows of $M$ are just a translation of the first one.

\section{Concatenation of tilings: extending perfect codes to larger dimensions}

In this section, we present some constructions to obtain new perfect codes out of a given pair of perfect codes. The principal tool to achieve the mentioned goal is the concatenation of tiles. We present here two main results. In Theorem \ref{teo:tilingconca} we consider concatenation of tiles that are balls of \textit{same} radius of two arbitrary TS-metrics and in Theorem \ref{teo:tilingposetconcat} we may consider balls of \emph{different} radii.

Since we are working with poly-tilings, the first step is to prove that the concatenation of poly-tilings results in a poly-tiling. That is what is stated in the next two results. The proof of  both will be omitted due to space limitations, but they follow directly from the definitions.
\vspace{2pt}

\begin{prop}\label{polyhe}
Let $D_1\subseteq \mathbb{F}_2^n$ and $D_2\subseteq\Fm$ and let  $D=D_1\mid D_2\subset \mathbb{F}_2^{n+m}$ be the concatenation of $D_1$ and $D_2$. Then, $D$ is a polyhedromino if, and only if, $D_1$ and $D_2$ are polyhedrominoes.
\end{prop}

In \cite[proof of Theorem 7.5]{vardy}, it was shown that given two tilings $(D_1,C_1)$ and $(D_2,C_2)$ of $\mathbb{F}_2^n$ and $\mathbb{F}_2^m$, respectively, the concatenation between $(D_1,C_1)$ and $(D_2,C_2)$ results in a tiling $(D,C)$ of $\mathbb{F}_2^{n+m}$. The same holds for poly-tilings. From this and Proposition \ref{polyhe}, we have the following:

\vspace{2pt}
\begin{coro}\label{concatenation}
	Let $(D_1,C_1)$ and $(D_2,C_2)$ be poly-tilings of $\mathbb{F}_2^n$ and $\mathbb{F}_2^m$, respectively. Then, $(D_1\mid D_2,C_1\mid C_2)   $ is a poly-tiling if, and only if, $(D_1,C_1)$ and $(D_2,C_2)$ are poly-tilings.

\end{coro}
\vspace{2pt}

Notice that the concatenation of two sets can be seen as a direct  product between them. Then, it would be natural to consider the product metric. But, in a general case, the concatenated tile $D$ is not  a metric ball in the product metric. For that reason, we define other metrics to accomplish our goal.   
From here on,  given $\bx\in\Fnm$, express $\bx:=\bx_1\mid \bx_2$, where $\bx_1\in\Fn$, $\bx_2\in\Fm$. %Notice that, in this work, we denote the vector $\bx$ in bold letter and $x_i$ as a coordinate of the vector $\bx$. But we adopt this new  
\vspace{2pt}
\begin{lema}\label{lema}
 Consider two metrics $d_1, d_2$ defined on $\Fn$ and $\Fm$ respectively and define  $d_{max}(\bx,\by):=max\{d_1(\bx_1,\by_1),d_2(\bx_2,\by_2)\}$. 
Then $d_{max}$ is a metric on  $ \Fnm$ and $d_1\in\mathcal{TS}(n)$, $d_2\in\mathcal{TS}(m)$  implies~$d_{max}\in\mathcal{TS}(m+n)$.
\end{lema}

The proof follows directly from the definition of a metric and it will be omitted due to lack of space. 

\vspace{2pt}
Now we consider the concatenation of two balls with same radius. 

\begin{teo}\label{teo:tilingconca}
	Let $(D_1,C_1), (D_2,C_2)$ be poly-tilings of $\mathbb{F}_2^n$ and $\mathbb{F}_2^m$, respectively. Suppose that $D_1=B_{d_{1}}({\bzero},r)$ and $D_2=B_{d_{2}}({\bzero},r)$, where $d_{1}, d_{2}$ are TS-metrics. 
		Let $(D,C) =(D_1\mid D_2, C_1\mid C_2) $. Then, $(D,C)$ is a poly-tiling of $\mathbb{F}_2^{n+m}$ and $D=B_{d_{max}}({\bzero},r)$.
\end{teo}

\vspace{2pt}
\begin{IEEEproof}
By Corollary \ref{concatenation} we have that $(D,C)$ is a poly-tiling. 
If $\bx\in D$ then $d_{max}(\bx,{\bzero})=max\{d_{1}(\bx_1,\bzero), d_{2}(\bx_2,\bzero)\}\leq r$, since $\bx_1\in D_1=B_{d_{1}}({\bzero},r)$ and $\bx_2\in D_2=B_{d_{2}}({\bzero},r)$. Thus, $\bx\in B_{d_{max}}({\bzero},r)$.
 If $\bx=\bx_1\mid \bx_2\notin D$ we have that $\bx_1\notin D_1$ or $\bx_2\notin D_2$, so that $d_{{1}}(\bx_1,\bzero)> r$ or $d_2(\bx_2,\bzero)>r$. But this implies that $d_{max}(\bx,\bzero)=max\{d_{1}(\bx_1,\bzero),d_{2}(\bx_2,\bzero)\}> r$ and $\bx\notin B_{d_{max}}({\bzero},r)$. 
Therefore, $D=B_{d_{max}}({\bzero},r)$.
\end{IEEEproof}

\vspace{2pt}
 In Theorem \ref{teo:tilingconca} we show that the concatenation $D=D_1\mid D_2$ of two TS-balls (which are poly-tilings) of same radius (possibly determined by different metrics) is a TS-ball. A natural question arises: is it possible to have different radii and $D$ be a ball? 
To answer this question we start constructing a TS-weight, made out of a conditional sum of weights.
 
% If yes, what are the conditions to be considered? Next lemma will provide a metric that turns $D$ a metric ball, considering $D_1$ and $D_2$ as balls with different radius.   

\begin{lema}\label{lema:rs}
		Let $\omega_1$ and $\omega_2$ be TS-weights on $\Fm$ and $\Fn$ respectively. Given $r\leq m,s\leq n$, let $D_1=B_{d_{1}}({\bzero},r)$,  $D_2=B_{d_{2}}({\bzero},s)$ and $D=D_1\mid D_2$, where $d_i$ is the metric determined by $\omega_i$. For $r\leq s$ we define the $s$-sum

\[\omega_1 \oplus_{s}^r   \omega_2(\bx) = 
\begin{cases}
\omega_1(\bx_1) +\omega_2(\bx_2), & \mbox{ if }  \bx\in D\\
r+s+1, & \mbox{ otherwise.} 
\end{cases}
\]
Then, $ \omega_1 \oplus_{s}^r   \omega_2 $ is a weight and it respects support.
\end{lema}

The proof follows directly from the definition of a weight and it will be omitted due to space limitations.
\vspace{2pt}

\begin{teo}\label{teo:tilingposetconcat}
Let $(D_1,C_1), (D_2,C_2)$ be TS-perfect codes. Then, $(D,C)=(D_1\mid D_2,C_1\mid C_2)$ is a TS-perfect code.
\end{teo}

\begin{IEEEproof}
	The hypothesis of the theorem ensures that  $D_1=B_{d_{1}}({\bzero},r)$ and $D_2=B_{d_{2}}({\bzero},s)$, where $d_1,d_2$ are TS-metrics, determined by weights $\omega_1,\omega_2$.  

 Corollary \ref{concatenation} ensures that $(D,C)$ is a poly-tiling of $\Fnm$. We assume $r\leq s$. Let  $\omega_1 \oplus_s^r   \omega_2$ be defined as in Lemma \ref{lema:rs}. From the lemma, all is left to prove is that $D=\{ \bx\in\Fnm ;  \omega_1 \oplus_s^r   \omega_2(\bx) \leq r+s  \}    $. 
From the definition of $\omega_1 \oplus_s^r   \omega_2$ we have that $\bx\in D$ if, and only if, $\omega_1 \oplus_s^r   \omega_2(\bx)\leq r+s$. 
\end{IEEEproof}

\section*{Acknowledgment}
Gabriella Akemi Miyamoto was supported by Capes (finance code 001) and CNPq. Marcelo Firer was partially supported by Sao Paulo Research Foundation, (FAPESP grant 2013/25977-7) and CNPq.

\bibliography{ref}{}
\bibliographystyle{plain}

\end{document}